\documentclass
[groupedaddress,doublelines,10pt,twocolumn,final,prl,showpacs]{revtex4}%
\usepackage{amssymb}
\usepackage{amsmath}
\usepackage{amsfonts}
\usepackage{graphicx}%
\providecommand{\U}[1]{\protect\rule{.1in}{.1in}}
\begin{document}
\title{Synopsis of a Unified Theory for All Forces and Matter}
\author{Zeng-Bing Chen}
\affiliation{National Laboratory of Solid State Microstructures, School of Physics, Nanjing
University, Nanjing 210093, China}
\date{\today }

\begin{abstract}
Assuming the Kaluza-Klein gravity interacting with elementary matter fermions
in a ($9+1$)-dimensional spacetime ($\mathcal{M}_{9+1}$), we propose an
information-complete unified theory for all forces and matter. Due to
entanglement-driven symmetry breaking, the $SO(9,1)$ symmetry of
$\mathcal{M}_{9+1}$ is broken to $SO(3,1)\times SO(6)$, where $SO(3,1)$
[$SO(6)$] is associated with gravity (gauge fields of matter fermions) in
$(3+1)$-dimensional spacetime ($\mathcal{M}_{3+1}$). The informational
completeness demands that matter fermions must appear in three families, each
having 16 independent matter fermions. Meanwhile, the fermion family space is
equipped with elementary $SO(3)$ gauge fields in $\mathcal{M}_{9+1}$, giving
rise to the Higgs mechanism in $\mathcal{M}_{3+1}$ through the gauge-Higgs
unification. After quantum compactification of six extra dimensions, a
trinity---the quantized gravity, the three-family fermions of total number 48,
and their $SO(6)$ and $SO(3)$ gauge fields---naturally arises in an effective
theory in $\mathcal{M}_{3+1}$. Possible routes of our theory to the Standard
Model are briefly discussed.

\end{abstract}

\maketitle

The tendency of unifying originally distinct physical subjects or phenomena
has profoundly advanced modern physics. Newton's law of universal gravitation,
Maxwell's theory of electromagnetism, and Einstein's relativity theory are
among the most outstanding examples for such a unification. The tremendous
successes of modern quantum field theory, i.e., the Standard Model (SM),
motivate the ambition of unifying all the four forces known so far---a kind of
\textquotedblleft theory of everything\textquotedblright. The superstring
theory and quantum gravity (particularly, loop quantum gravity---LQG
\cite{Rovelli-book,Thiemann,loop25,Chiou}), both with remarkable results, are
two tentative proposals. Here we take a more \textquotedblleft
orthodox\textquotedblright\ viewpoint following the SM and LQG, rather than
the superstring theory.

The tradition of physics, initiated from Newton, is to describe a physical
system by dynamical laws, usually in terms of differential equations. To
determine the (classical or quantum) state of the system, the initial (as well
as boundary) conditions have to be given regardless of dynamical laws.
Different allowed initial conditions lead to different solutions to dynamical
laws. Such a tradition is called Newton's paradigm \cite{Smolin-PT}, which is
believed not to apply to the whole Universe. For the unique Universe the
initial conditions must themselves be a part of physical laws. Thus, if any
form of the theory of everything is conceivable at all, it must unify
dynamical laws and states, i.e., break down the distinction between dynamical
laws and initial conditions \cite{Smolin-unify} such that it applies to the
Universe as a whole.

Recently, we attempted another unification by unifying spacetime and matter as
information via an information-complete quantum field theory, which describes
elementary fermions, their gauge fields and spacetime (gravity) as a trinity
\cite{ICQT,icQFT}. Therein, complete physical information of the trinity is
encoded in dual entanglement---spacetime-matter entanglement and
\textquotedblleft programmed\textquotedblright\ entanglement between
elementary fermions and their gauge fields (together as matter), a fact called
the information-completeness principle (ICP). The basic state-dynamics
postulate \cite{icQFT} is that the Universe is self-created into a state
$\left\vert e,\omega;A...,\psi...\right\rangle $ of all physical contents
(spacetime and matter), from no spacetime and no matter, with the least action
($\hbar=c=1$)
\begin{align}
&  \left.  \left\vert e,\omega;A...,\psi...\right\rangle =e^{iS_{\mathrm{GM}%
}(e,\omega;A...,\psi...)}\left\vert \emptyset\right\rangle ,\right.
\nonumber\\
&  \left.  \delta S_{\mathrm{GM}}(e,\omega;A...,\psi...)\left\vert
e,\omega;A...,\psi...\right\rangle =0.\right.  \label{postuD}%
\end{align}
Here $\left\vert \emptyset\right\rangle \equiv\left\vert \emptyset
_{\mathrm{G}}\right\rangle \otimes\left\vert \emptyset_{\mathrm{M}%
}\right\rangle $ is the common vacuum state of matter (the matter vacuum
$\left\vert \emptyset_{M}\right\rangle $) and geometry (the empty-geometry
state $\left\vert \emptyset_{G}\right\rangle $ in LQG \cite{Rovelli-book});
$A...$ ($\psi...$) represent all gauge (matter fermions) fields; gravity is
described by the tetrad field $e_{\mu}^{a}$ and the spin connection
$\omega_{\mu}^{ab}$, to be specified below. Note that the dynamical law and
states always appear jointly in the postulate [Eq.~(\ref{postuD})]. This is in
sharp contrast to the tradition where the dynamical law [i.e., $\delta
S_{\mathrm{GM}}(e,\omega;A...,\psi...)=0$] and states are given separately.
Thus, besides the conceptual advantages as shown previously \cite{icQFT}, the
theory unifies the dynamical law and states.

The conceptual advantages of the information-complete quantum description
motivate us to consider the ultimate unification of all known forces in this
Letter. As is well-known, consistent superstring theory (or its updated
version, the $M$-theory) exists only in ($9+1$)- or ($10+1$)-dimensional
spacetime. Here we assume a ($9+1$)-dimensional spacetime ($\mathcal{M}_{9+1}%
$) instead of the usual ($3+1$)-dimensional one ($\mathcal{M}_{3+1}$), but do
not assume string and supersymmetry. Then the trinary fields of our theory
consist of the ($9+1$)-dimensional Kaluza-Klein gravity \cite{KK-rev}
interacting with elementary matter fermions and an elementary gauge field. The
ICP predicts that (1) matter fermions must appear in three families, each of
which has 16 independent fermions, and (2) there are the $SO(3)$ family gauge
fields. A new, entanglement-induced symmetry breaking mechanism and
compactifying six extra dimensions then lead to the ($3+1$)-dimensional
trinity, in which the Higgs fields naturally arise.

\textit{Information-complete unification}.---Matter fermions in the SM are six
quarks ($u$, $d$; $c$, $s$; $t$, $b$) and six leptons (electron $e$, electron
neutrino $\nu_{e}$; muon $\mu$, muon neutrino $\nu_{\mu}$; tau $\tau$, tau
neutrino $\upsilon_{\tau}$). They can be precisely grouped into three
\textquotedblleft families\textquotedblright\ (or \textquotedblleft
generations\textquotedblright) as ($\upsilon_{e}$, $e$, $u$, $d$),
($\upsilon_{\mu}$, $\mu$, $c$, $s$), and ($\upsilon_{\tau}$, $\tau$, $t$,
$b$). Together with the corresponding antiparticles, totally we have $45$
matter fermions if each neutrino is merely left-handed. The three families
have identical properties, except for distinct mass patterns. The origin of
this amazing structure is a long-standing puzzle in the SM, which, together
with the mass-generating Higgs mechanism, describes very successfully these
fermions interacting via $SU(3)\times SU(2)\times U(1)$ gauge fields (the
strong, weak, and electromagnetic forces). The particular SM group structure
was discovered empirically; \textit{there is no fundamental principle
dicatating why we should choose this particular group, but not others}
\cite{Kane-book}. Some Grand Unification Theories (GUTs) extended the SM group
into larger groups, such as $SU(5)$ \cite{su5} and $SO(10)$ \cite{so10-g,so10}.

Here we assume a ($9+1$)-dimensional spacetime (a curved manifold
$\mathcal{M}_{9+1}$) with an $SO(9,1)$ symmetry, where the coordinates
$x=(x^{A})=(x^{\mu},x^{\bar{\mu}})$ with \textquotedblleft curved
indices\textquotedblright\ $\mu=0,1,2,3$ and $\bar{\mu}=4,5,...,9$. A
Minkowski vector is denoted by $y=(y^{I})=(y^{a},y^{\bar{a}})$ with
\textquotedblleft flat indices\textquotedblright\ $a=0,1,2,3$ and $\bar
{a}=4,5,...,9$; the Minkowski metric $\eta_{IJ}$ has signature $\left[
-,+,...,+\right]  $. Gravity in $\mathcal{M}_{9+1}$ is then described by the
tetrad field $\hat{e}_{A}^{I}(x)$ (with the inverse $\hat{e}_{I}^{A}$), which
relates the ($9+1$)-dimensional metric $\bar{g}_{AB}(x)$ via $\bar{g}%
_{AB}(x)=\hat{e}_{A}^{I}(x)\hat{e}_{B}^{J}(x)\eta_{IJ}$. As the $SO(9,1)$
group has 45 generators, gravity in $\mathcal{M}_{9+1}$ has 45 independent
field components and thus 90 independent internal states provided that the
polarization degree of freedom (DoF) is taken in account.

Matter in $\mathcal{M}_{9+1}$ is assumed to be an elementary fermion field
$\psi$ in the spinorial representation of $SO(9,1)$, which has $2^{5}=32$
dimensions. In terms of the Dirac matrices $\gamma_{a}$ and $\gamma_{5}$ for
$\mathcal{M}_{3+1}$ and $\tilde{\gamma}_{\bar{a}}$ for six extra dimensions
($\mathcal{M}_{6}$), one can construct $10$ $\Gamma$-matrices \cite{KK-rev}
\begin{equation}
\Gamma_{a}=\gamma_{a}\otimes I,\text{\quad\quad}\Gamma_{\bar{a}}=i\gamma
_{5}\otimes\tilde{\gamma}_{\bar{a}-3},\label{dm}%
\end{equation}
which satisfy $\{\Gamma_{I},\Gamma_{J}\}=\Gamma_{I}\Gamma_{J}+\Gamma_{J}%
\Gamma_{I}=2\eta_{IJ}I$. These $\Gamma$-matrices then form the spinorial
representation of $SO(9,1)$ as
\begin{align}
\lbrack\Pi_{IJ},\Pi_{KL}] &  =i(\eta_{IL}\Pi_{JK}+\eta_{JK}\Pi_{IL}\nonumber\\
&  -\eta_{IK}\Pi_{JL}-\eta_{JL}\Pi_{IK}),\label{so10}%
\end{align}
with $45$ generators $\Pi_{IJ}=\frac{i}{4}[\Gamma_{I},\Gamma_{J}]$. The
32-dimensional representation implies that the matter fermions would have 32
independent internal states if no further constraint is required. Recall that
the SM describes chiral fermions such that fermions of different chiralities
transform differently under $SU(3)\times SU(2)\times U(1)$. In particular,
each SM fermion family has only 30 internal states (i.e., 15 independent
matter fermions).

Let us consider the input arising from our information-completeness trinary
description. According to the general formalism of an information-complete
quantum field theory \cite{icQFT}, all physical predictions are encoded by
spacetime-matter entanglement, which can be decomposed in a Schmidt form such
that spacetime (i.e., gravity) and matter are mutually defined to acquire
information. For our description to be information-complete, the number of
independent gravity field components and the number of independent gauge field
components (if any) must match the number of independent matter fermions. As
the spinorial representation of $SO(9,1)$ has only 16 elementary fermions,
\textit{there must be three and only three fermion families (duplicates), each
of which belongs to the spinorial representation of }$SO(9,1)$. Hereafter, we
then denote the matter fermion fields by $\psi_{\alpha}$ with the family
indices $\alpha=1,2,3$. Then the number of independent gauge field components
is determined to be $16\times3-45=3$. Note that the family indices label a new
internal space of fermions, or a new quantum number, which is physically
different from the quantum number characterizing fermions of each family. If
we assume that the (elementary) gauge field acts on this family space (the
family triplet), it is natural to require the family gauge fields to be
$SO(3)$ gauged, namely, the family triplet of the matter fermions form a basis
for a 3-dimensional irreducible representation of $SO(3)$: $[t_{\alpha
},t_{\beta}]=i\varepsilon_{\alpha\beta\gamma}t_{\gamma}$. Here $t_{\alpha}$
are the $3$ generators of $SO(3)$, $(t_{\alpha})_{\beta\gamma}=-i\varepsilon
_{\alpha\beta\gamma}$ for the adjoint representation, and the $SO(3)$
structure constants $\varepsilon_{\alpha\beta\gamma}$\ are totally
antisymmetric and $\varepsilon_{123}=1$.

Now it is ready to write down the total action of gravity and matter in
$\mathcal{M}_{9+1}$. We can use the beautiful language of differential forms
to express the relevant geometry. For instance, $\hat{e}^{I}(x)=\hat{e}%
_{A}^{I}(x)dx^{A}$ represents 1-form. The infinitesimal rotation of $\hat
{e}^{I}$ is then a 2-form $d\hat{e}^{I}=-\hat{\omega}_{\text{ }J}^{I}%
\wedge\hat{e}^{J}$ (Note that $d\hat{e}^{I}+\hat{\omega}_{\text{ }J}^{I}%
\wedge\hat{e}^{J}\equiv\hat{T}^{I}$ is the torsion; here we consider therefore
a torsion-free manifold), where an antisymmetric 1-form $\hat{\omega}%
^{IJ}=-\hat{\omega}^{JI}$ is the so-called spin connection. With the
connection 1-form, a tensor 2-form $\hat{R}^{IJ}=d\hat{\omega}^{IJ}%
+\hat{\omega}_{\text{ }K}^{I}\wedge\hat{\omega}^{KJ}=\hat{R}_{AB}^{IJ}%
dx^{A}dx^{B}$ can be defined and is known as the curvature. The scalar
curvature reads $\hat{R}=\hat{R}_{AB}^{IJ}\hat{e}_{I}^{A}\hat{e}_{J}^{B}$. The
total action then reads $\hat{S}_{\mathrm{GM}}^{(9+1)}(\hat{e},\hat{\omega
};\mathbf{\psi};\mathbf{z})=\hat{S}_{\mathrm{G}}^{(9+1)}(\hat{e},\hat{\omega
})+\hat{S}_{\mathrm{G+M}}^{(9+1)}(\hat{e},\hat{\omega};\mathbf{\psi
};\mathbf{z})=\hat{S}_{\mathrm{G}}^{(9+1)}+\hat{S}_{\mathrm{G+g}}^{(9+1)}%
+\hat{S}_{\mathrm{G+D+g}}^{(9+1)}$, where
\begin{align}
\hat{S}_{\mathrm{G}}^{(9+1)}  &  =\frac{1}{16\pi G_{10}}\int_{\mathcal{M}%
_{9+1}}dx^{10}\hat{e}(\hat{R}+\Lambda_{10}),\nonumber\\
\hat{S}_{\mathrm{G+g}}^{(9+1)}  &  =-\frac{1}{4}\int_{\mathcal{M}_{9+1}%
}dx^{10}\hat{e}f_{AB}^{\alpha}f^{AB,\alpha},\nonumber\\
\hat{S}_{\mathrm{G+D+g}}^{(9+1)}  &  =\int_{\mathcal{M}_{9+1}}dx^{10}\hat
{e}\bar{\psi}_{\beta}\Gamma^{I}\hat{e}_{I}^{A}iD_{A}^{\beta\gamma}\psi
_{\gamma}. \label{Sgm}%
\end{align}
Here $G_{10}$ ($\Lambda_{10}$) represents the Newton (cosmological) constant
in $\mathcal{M}_{9+1}$, $\hat{e}=\left\vert \det\hat{e}_{I}^{A}\right\vert $;
$\mathbf{\psi}=(\psi_{\alpha})$, $\mathbf{\bar{\psi}}=\mathbf{\psi}^{\dag
}\Gamma^{0}$, the covariant derivative of Dirac's spinors reads $D_{A}%
^{\beta\gamma}=\delta_{\beta\gamma}\partial_{A}+i\delta_{\beta\gamma}%
\hat{\omega}_{A}^{IJ}\Pi_{IJ}-ig_{3}(t_{\alpha})_{\beta\gamma}z_{A}^{\alpha}$,
where the field strengths of the $SO(3)$ family gauge fields are
$f_{AB}^{\alpha}=\partial_{A}z_{B}^{\alpha}-\partial_{B}z_{A}^{\alpha}%
+g_{3}\varepsilon_{\alpha\beta\gamma}z_{A}^{\beta}z_{B}^{\gamma}$ with
coupling constant $g_{3}$.

The theory is not a pure Kaluza-Klein theory, but \textit{requires/predicts
the existence} of the explicit $SO(3)$ family gauge fields even in
$\mathcal{M}_{9+1}$, showing the \textit{necessity of our trinary description
at the most fundamental level}. The presence of the elementary gauge fields is
advantageous \cite{KK-rev} to compactify the extra dimensions, to provide an
origin of the Higgs fields, and to avoid the difficulty of obtaining chiral
fermions in $\mathcal{M}_{3+1}$ for pure Kaluza-Klein theories (For chirality
in $\mathcal{M}_{9+1}$, see also \cite{Witten}). The state and dynamics of the
elementary trinity are then given by
\begin{align}
&  \left.  \left\vert \hat{e},\hat{\omega};\mathbf{\psi};\mathbf{z}%
\right\rangle =e^{i\hat{S}_{\mathrm{GM}}^{(9+1)}(\hat{e},\hat{\omega
};\mathbf{\psi};\mathbf{z})}\left\vert \emptyset\right\rangle ,\right.
\nonumber\\
&  \left.  \delta\hat{S}_{\mathrm{GM}}^{(9+1)}(\hat{e},\hat{\omega
};\mathbf{\psi};\mathbf{z})\left\vert \hat{e},\hat{\omega};\mathbf{\psi
};\mathbf{z}\right\rangle =0.\right.  \label{sd10}%
\end{align}
The above considerations fix our information-complete GUT with gravity\ (or
the Grand Integration Theory, GIT). Now it is amazing to see that our theory
is \textit{uniquely} determined by the spacetime symmetry, the gauge
principle, \textit{and} the ICP. The GIT, however, does not assume a single
gauge group as in the usual GUT. Here the physical predictions of the GIT are
the dual entanglement \cite{ICQT,icQFT} in $\left\vert \hat{e},\hat{\omega
};\mathbf{\psi};\mathbf{z}\right\rangle $, i.e., entanglement between
($9+1$)-dimensional Kaluza-Klein gravity and matter, as well as entanglement,
programmed by gravity, between the three-family fermions and the $SO(3)$
family gauge fields. The second line of Eq.~(\ref{sd10}) gives the equations
of motion, conservation laws, and constraints of the GIT, all acting on
$\left\vert \hat{e},\hat{\omega};\mathbf{\psi};\mathbf{z}\right\rangle $.
Below we consider the effective theory of the GIT in $\mathcal{M}_{3+1}$.

\textit{Entanglement-driven symmetry breaking}.---Now let us consider the
physical consequences of gravity-matter entanglement. After gravity-matter
entangling, the original internal space of the Kaluza-Klein gravity will be
physically differentiated by the family quantum number and as such, the
original $SO(9,1)$ symmetry of the Kaluza-Klein gravity must be broken to a
lower symmetry. This new mechanism of symmetry breaking can thus be called
entanglement-driven symmetry breaking.

Mathematically, the maximal subgroup of $SO(9,1)$ is $SO(3,1)\times SO(6)$
which has the same rank (namely, 5) as $SO(9,1)$.\ The most obvious way is to
choose the lower symmetry to be $SO(3,1)\times SO(6)$, resulting in the
conventional $\mathcal{M}_{3+1}$ (or the usual gravity) and $6$-dimensional
compactified space $\mathcal{M}_{6}$; other choices lead to inconsistencies
this way or another by following the procedure given below. As a result of
symmetry breaking from $SO(9,1)$ to $SO(3,1)\times SO(6)$, $\psi_{\alpha}$ for
given family should be the spinorial representations of $SO(3,1)$ and of
$SO(6)$ \textit{simultaneously}, which are $4$- and $8$-dimensional,
respectively. The total dimensions of $\psi_{\alpha}$ are again 32. Meanwhile,
symmetry breaking singles out $\mathcal{M}_{3+1}$ gravity as the
\textquotedblleft programming system\textquotedblright\ \cite{ICQT,icQFT}.

Then, according to the usual Kaluza-Klein mechanism, one can associate the
Lorentz group $SO(3,1)$ with gravity in $\mathcal{M}_{3+1}$ and $SO(6)$ with
the gauge field of matter fermions. Namely, we can expand $\bar{g}_{AB}(x)$
with the following ansatz \cite{KK-rev,Cho}
\begin{equation}
\bar{g}_{AB}=\left(
\begin{array}
[c]{cc}%
g_{\mu\nu}(x^{\mu})+\tilde{g}_{\bar{\mu}\bar{\nu}}(x^{\bar{\mu}})W_{\mu}%
^{\bar{\mu}}W_{\nu}^{\bar{\nu}} & W_{\mu}^{\bar{\nu}}\\
W_{\nu}^{\bar{\mu}} & \tilde{g}_{\bar{\mu}\bar{\nu}}(x^{\bar{\mu}})
\end{array}
\right)  \label{gab}%
\end{equation}
with $W_{\mu}^{\bar{\nu}}=\xi_{\bar{a}\bar{b}}^{\bar{\nu}}(x^{\bar{\mu}%
})\mathbf{Z}_{\mu}^{\bar{a}\bar{b}}(x^{\mu})$, where $\xi_{\bar{a}\bar{b}%
}^{\bar{\nu}}(x^{\bar{\mu}})$ is the Killing vectors appearing in the
infinitesimal isometry $I+i\varepsilon^{\bar{a}\bar{b}}\Sigma_{\bar{a}\bar{b}%
}:x^{\bar{\nu}}\rightarrow x^{\prime\bar{\nu}}=x^{\bar{\nu}}+\varepsilon
^{\bar{a}\bar{b}}(x^{\mu})\xi_{\bar{a}\bar{b}}^{\bar{\nu}}(x^{\bar{\mu}})$;
the infinitesimal parameters $\varepsilon^{\bar{a}\bar{b}}(x^{\mu})$ are
independent of $x^{\bar{\mu}}$. The ansatz (\ref{gab}), after integrating the
extra dimensions in $\hat{S}_{\mathrm{GM}}^{(9+1)}(\hat{e},\hat{\omega
};\mathbf{\psi};\mathbf{z})$, results in the $(3+1)$-dimensional actions for
gravity and the $SO(6)$ gauge field $\mathbf{Z}_{\mu}^{\bar{a}\bar{b}}$ (with
coupling constant $g_{6}$)
\begin{align}
S_{\mathrm{G}}^{\bar{\Lambda}}  &  =\frac{1}{16\pi G}\int_{\mathcal{M}_{3+1}%
}dx^{4}e(R+\bar{\Lambda}),\nonumber\\
S_{\mathrm{gG}}  &  =-\frac{1}{4}\int_{\mathcal{M}_{3+1}}dx^{4}e\mathbf{F}%
_{\mu\nu}^{\bar{a}\bar{b}}\mathbf{F}^{\mu\nu,\bar{a}\bar{b}}. \label{Ggauge}%
\end{align}
Here $e=\left\vert \det e_{a}^{\mu}\right\vert =\sqrt{-\det g_{\mu\nu}}$, $R$
stands for the scalar curvature in $\mathcal{M}_{3+1}$, $G$ is the usual
Newton constant, and the cosmological term $\bar{\Lambda}$ gathers also the
effects of compactified space on gravity in $\mathcal{M}_{3+1}$\ \cite{Cho};
the $SO(6)$ field strengths $\mathbf{F}_{\mu\nu}^{\bar{a}\bar{b}}%
=\partial_{\mu}\mathbf{Z}_{\nu}^{\bar{a}\bar{b}}-\partial_{\nu}\mathbf{Z}%
_{\mu}^{\bar{a}\bar{b}}+g_{6}(\mathbf{Z}_{\mu}^{\bar{a}\bar{c}}\mathbf{Z}%
_{\nu}^{\bar{b}\bar{c}}-\mathbf{Z}_{\nu}^{\bar{a}\bar{c}}\mathbf{Z}_{\mu
}^{\bar{b}\bar{c}})$. Thus, by treating $\mathcal{M}_{6}$ and $\mathcal{M}%
_{3+1}$ on an equal footing, non-Abelian gauge fields naturally arise from a
higher-dimensional gravity.

The $\mathcal{M}_{3+1}$ action for the family gauge fields reads
\begin{equation}
S_{\mathrm{fG}}=-\frac{1}{4}\int_{\mathcal{M}_{3+1}}dx^{4}ef_{\mu\nu}^{\alpha
}f^{\mu\nu,\alpha}. \label{famg}%
\end{equation}
Following the idea of gauge-Higgs unification proposed long ago
\cite{Manton,Manton-chiral}, the extra components of $z_{A}^{\alpha}$ living
in $\mathcal{M}_{6}$ are interpreted as the Higgs fields $\phi_{\alpha}$. We
denote the action of the Higgs fields interecting with $z_{\mu}^{\alpha}$ and
gravity by $S_{\mathrm{HfG}}$. The action of the Dirac sector (matter fermions
interacting with gravity, the gauge fields, and $\phi_{\alpha}$) is
\begin{equation}
S_{\mathrm{DHgfG}}=S_{\mathrm{DHG}}+S_{\mathrm{DgfG}}, \label{gdg}%
\end{equation}
where $S_{\mathrm{DgfG}}=\int_{\mathcal{M}_{3+1}}dx^{4}e\bar{\psi}_{\beta
}\gamma^{a}e_{a}^{\mu}iD_{\mu}^{\beta\gamma}\psi_{\gamma}$ with $D_{\mu
}^{\beta\gamma}=\delta_{\beta\gamma}(\partial_{\mu}+i\omega_{\mu}^{ab}\Pi
_{ab}-ig\mathbf{Z}_{\mu}^{\bar{a}\bar{b}}\Pi_{\bar{a}\bar{b}})-ig_{3}%
(t_{\alpha})_{\beta\gamma}z_{\mu}^{\alpha}$; $S_{\mathrm{DHG}}$ is the action
for the Higgs fields coupling with the Dirac fields. Here we do not give the
explicit forms of $S_{\mathrm{HfG}}$ and $S_{\mathrm{DHG}}$ and leave them for
a future work.

\textit{Quantum compactification\ of extra dimensions}.---Usually,
compactification of extra dimensions in the Kaluza-Klein and superstring
theories is a complicated and unfinished issue. Below we further show that we
can insist on the information-complete trinary description even in
$\mathcal{M}_{3+1}$ as a robust theoretical structure and as such, we could
vastly simplify the problem of uncovering the reliable feature of the
compactification without going into its detailed physics.

Let us first focus on $\mathcal{M}_{3+1}$. It is well-known that the Lorentz
algebra $SO(3,1)$ is locally isomorphic to $SU(2)\times SU(2)$. Here one
$SU(2)$\ generates the rotations and another the boosts. In LQG
\cite{Rovelli-book,Thiemann,loop25,Chiou}, gravity in $\mathcal{M}_{3+1}$
possesses only one $SU(2)$ gauge structure related to the rotations, while the
boost DoFs are not dynamical \cite{Cianfrani}. After the $3+1$ spacetime
decomposition and taking the time gauge, one arrives at the Hamiltonian
formalism \cite{Rovelli-book,Chiou}. The dynamical variables of the $SU(2)$
gravity, in terms of $e_{a}^{\mu}$ and $\omega_{\mu}^{ab}$, are the connection
field $\mathcal{\hat{A}}_{m}^{r}$ [defined on $\mathcal{M}_{3}$; $m$: spatial
indices and $r$: $SU(2)$-valued], whose conjugate variable is the
\textquotedblleft gravitational electric field\textquotedblright%
\ $\mathcal{\hat{E}}_{n}^{s}$. The canonical dynamical variables for
$\mathbf{Z}_{\mu}^{\bar{a}\bar{b}}$, $z_{\mu}^{\alpha}$, and $\psi_{\alpha}$,
whose explicit forms are not important in subsequent discussions, can also be
obtained within LQG. A remarkable result of LQG is to identify the state space
of quantized gravity, spanned by a complete orthogonal basis $\left\{
\left\vert \Gamma,\{j_{l}\},\{i_{n}\}\right\rangle \right\}  $ consisted of
the spin-network states with respect to an abstract graph $\Gamma$ (with nodes
$n$ and oriented links $l$) in three-dimensional region $\mathcal{R}$ with
boundary $\partial\mathcal{R}$. Here $j_{l}$ is an irreducible $j$
representation of $SU(2)$ for each link $l$ and $i_{n}$ the $SU(2)$
intertwiner for each node $n$.

As we discussed above, there are three families of matter fermions. Quantum
mechanically, the family information can be encoded by introducing three
internal states $\left\vert \mathrm{f}_{\alpha}\right\rangle $ attached on the
fermion state space; \{$\left\vert \mathrm{f}_{\alpha}\right\rangle $\} forms
a complete orthogonal basis. Then according to the ICP, we have to introduce
another quantum system interacting/entangling with the fermion family space
such that the family information can be acquired. Such a quantum system can
only be attributed to the compactified space, which must be effectively a
simple three-state system. The three (the number is required by the ICP to be
the same as the family number) states are denoted by $\left\vert
\mathcal{\alpha},\mathrm{6D}\right\rangle \equiv\left\vert \mathcal{\alpha
}\right\rangle $ ($\alpha=1,2,3$). Of course, all physical DoFs related to
$\mathcal{M}_{6}$ are quantized, resulting in a complicated state. Yet, in the
\textquotedblleft quantum compactification\textquotedblright\ mechanism
proposed here, the compactified $\mathcal{M}_{6}$ has only three states
accessible to physical DoFs in $\mathcal{M}_{3+1}$.

After such a compactification, all the remaining forces fields and the fermion
fields are confined in $\mathcal{M}_{3+1}$ such that their field quanta are
massless as protected by the relevant symmetries. Meanwhile, each $\left\vert
\mathcal{\alpha}\right\rangle $ encodes all information for matter and gravity
(including the gauge fields with mixed indices of $\mathcal{M}_{3+1}$ and
$\mathcal{M}_{6}$) related to $\mathcal{M}_{6}$ although the quantized
$\mathcal{M}_{6}$ itself is merely an effective three-state system. This
immediately means that information encoded in $\left\vert \mathcal{\alpha
}\right\rangle $ (except for the three states), as well as the related matter,
is \textquotedblleft dark\textquotedblright\ with respect to the gauge fields
in $\mathcal{M}_{3+1}$. Thus, our GIT enables a quantum information definition
of (non-Abelian) dark matter, which is dark to non-Abelian gauge fields
survived in $\mathcal{M}_{3+1}$; for dark energy, see \cite{icQFT}.

In terms of the spin-network basis, following the formalism in
Ref.~\cite{icQFT} we propose the quantum-compactification ansatz for quantized
spacetime $\mathcal{M}_{9+1}$
\begin{equation}
\sum_{\substack{l\in\Gamma\cap\partial\mathcal{R}\\n\in\Gamma\cap
\mathcal{R},\alpha}}\mathrm{S}_{\Gamma}(l,n,\alpha)\left\vert \Gamma
,j_{l},i_{n};\alpha\right\rangle \left\vert \mathcal{\alpha}\right\rangle
\equiv\sum_{\alpha}\mathrm{s}_{\Gamma}(\alpha)\left\vert \Gamma,\alpha
\right\rangle \left\vert \mathcal{\alpha}\right\rangle ,\label{ansatz}%
\end{equation}
where three $\left\vert \Gamma,\alpha\right\rangle $ form an orthogonal basis
and are the \textquotedblleft three-world\textquotedblright\ states
corresponding to $\left\vert \mathcal{\alpha}\right\rangle $ and the three
families of matter fermions. Entanglement between quantized spacetime
$\mathcal{M}_{3+1}$ and $\mathcal{M}_{6}$, as shown in Eq.~(\ref{ansatz}),
suggests interaction between them. To be consistent with the Einstein
equations in $\mathcal{M}_{3+1}$, the only possible form of the interaction
would be
\begin{equation}
S_{\mathrm{G,6D}}=\frac{1}{16\pi G}\sum_{\alpha}\int_{\mathcal{M}_{3+1}}%
dx^{4}e\tilde{\Lambda}_{\alpha}\left\vert \mathcal{\alpha}\right\rangle
\left\langle \alpha\right\vert .\label{cos6d}%
\end{equation}
Namely, our GIT predicts the cosmological constant $\Lambda\equiv\bar{\Lambda
}+\sum_{\alpha}\tilde{\Lambda}_{\alpha}\left\vert \mathcal{\alpha
}\right\rangle \left\langle \alpha\right\vert $\ to be an operator in
\{$\left\vert \mathcal{\alpha}\right\rangle $\}; the total cosmological term
contains the remnant effect of the compactified space. Similarly to the ansatz
in Eq.~(\ref{ansatz}), we conjecture that there is also interaction
(entanglement) between the fermion family space and the compactified space
[see Eq.~(\ref{UnivS}) below]. The corresponding action is denoted by
$S_{\mathrm{DG,6D}}$, whose explicit form is a future work.

\textit{Total action and dynamics}.---To sum up the above results, the total
action in $\mathcal{M}_{3+1}$ is
\begin{equation}
S_{\mathrm{GIT}}=S_{\mathrm{G}}^{\Lambda}+S_{\mathrm{gG}}+S_{\mathrm{fG}%
}+S_{\mathrm{DHgfG}}+S_{\mathrm{HfG}}+S_{\mathrm{DG,6D}}, \label{totalS}%
\end{equation}
with $S_{\mathrm{G}}^{\Lambda}=S_{\mathrm{G}}^{\bar{\Lambda}}+S_{\mathrm{G,6D}%
}$. While we leave physics of the Higgs mechanism for future work, it should
be noted that in our GIT, each family of matter fermions is labeled by one
Higgs field and entangled with one of the three family gauge fields after
symmetry breaking of the family symmetry; the $SO(6)$ gauge fields act
universally upon the three families. Thus, for each family the total number of
gauge fields is $15+1$, matching the number of elementary fermions. This shows
that our GIT is information-complete even in $\mathcal{M}_{3+1}$. Below we
consider the dynamics of the physical Universe described by $S_{\mathrm{GIT}}$.

The information-complete trinary description allows us to adopt either a
\textquotedblleft global view\textquotedblright\ or a \textquotedblleft local
view\textquotedblright\ of the dynamics. The global view is reflected by the
fact that all physical predictions about the whole Universe are encoded in
gravity-matter entanglement $\left\vert \mathcal{A},\alpha,(\mathbf{\psi
},\mathbf{z},\mathbf{Z})\right\rangle =e^{iS_{\mathrm{GIT}}}\left\vert
\emptyset\right\rangle \otimes\left\vert \alpha_{0}\right\rangle $ for given
\textquotedblleft initial state\textquotedblright\ $\left\vert \alpha
_{0}\right\rangle $ of $\mathcal{M}_{6}$. Thanks to the spin-network basis,
\begin{align}
\left\vert \mathcal{A},\{\alpha\},(\mathbf{\psi},\phi,\mathbf{z}%
,\mathbf{Z})\right\rangle  &  =\sum_{\substack{l\in\Gamma\cap\partial
\mathcal{R}\\n\in\Gamma\cap\mathcal{R},\alpha}}\mathrm{S}_{\Gamma}%
(l,n,\alpha)\left\vert \Gamma,j_{l},i_{n};\alpha;t\right\rangle \nonumber\\
&  \otimes\left\vert \mathcal{\alpha}\right\rangle \otimes\left\vert
(\psi,\phi,\mathbf{z},\mathbf{Z});\Gamma,l,n;\alpha;t\right\rangle .
\label{UnivS}%
\end{align}
Here $t$ denotes time; $\mathrm{S}_{\Gamma}$ must be time-independent as
$\left\vert \mathcal{A},\{\alpha\},(\mathbf{\psi},\phi,\mathbf{z}%
,\mathbf{Z})\right\rangle $ is annihilated by, among others, the total
Hamiltonian,\ known as the Hamiltonian constraint. Because a spin-network
state $\left\vert \Gamma,j_{l},i_{n};\alpha;t\right\rangle $ defines spacetime
and thus, \textit{is} spacetime \cite{Rovelli-book}, we can include $t$
explicitly in $\left\vert \Gamma,j_{l},i_{n};\alpha;t\right\rangle $.
$\left\vert (\psi,\phi,\mathbf{z},\mathbf{Z});\Gamma,l,n;\alpha;t\right\rangle
\equiv\left\vert (\psi,\phi,\mathbf{z},\mathbf{Z});\Gamma,k_{l},F_{n}%
,S_{n},w_{n};\alpha;t\right\rangle $ ($F_{n},S_{n},w_{n}$: number of fermions,
number of Higgs scalars, and the field strength at node $n$, respectively,
$k_{l}$: flux of the electric gauge fields\ across surface $l$; see
Refs.~\cite{Rovelli-book,Chiou}), programmed by a given spin-network state and
by $\left\vert \mathcal{\alpha}\right\rangle $, encodes entanglement between
(i.e., all physical predictions about) matter fermions and their gauge fields
(the Higgs scalars are the extra components of the family gauge fields). Such
a particular entanglement structure of the Universe is called dual
entanglement \cite{ICQT,icQFT}. Following Ref.~\cite{icQFT}, the evolution
operator $U_{\mathrm{GM}}\equiv\exp[iS_{\mathrm{GIT}}]$ has a factorizable
structure. In this way, the dynamics of the Universe can be cast into a dual
form without the notorious \textquotedblleft problem of time\textquotedblright%
\ \cite{Rovelli-book,Chiou} in quantum gravity, thus recovering a description
of the local view.

Note that all states for gravity, matter fermions, and gauge fields, once
Schmidt-decomposed in the dual form, are physical predictions of the theory
and thus must be annihilated automatically by any constraints appearing in the
theory. This fact embodies again that our formalism unifies the dynamical law
and states.

\textit{Possible routes to the SM}.---To show more predictive power of the
GIT, let us briefly consider chirality, which is a remarkable feature of the
SM. For pure Kaluza-Klein theories, it is difficult to obtain chiral fermions
\cite{KK-rev,Witten}. But in our theory, the situation is dramatically
different. In $\mathcal{M}_{9+1}$, the chirality operator $\chi=i\gamma
_{5}\otimes\Gamma_{\mathrm{6D}}^{\mathrm{five}}$, where $\gamma_{5}$
($\Gamma_{\mathrm{6D}}^{\mathrm{five}}\equiv\tilde{\gamma}_{1}\tilde{\gamma
}_{2}...\tilde{\gamma}_{6}$) is the chirality for $\mathcal{M}_{3+1}$
($\mathcal{M}_{6}$); the $\mathcal{M}_{3+1}$ chirality and the $\mathcal{M}%
_{6}$ chirality are correlated \cite{KK-rev,Manton-chiral,Witten}. Now if the
elementary $SO(3)$ [or its isomorphic group, $SU(2)$] gauge fields have
topologically non-trivial background (e.g., the monopole or instanton
configuration) in $\mathcal{M}_{6}$, one can have chiral zero modes of
fermions forming a complex representation of the symmetry group
\cite{KK-rev,Salam,origin,Hosotani}. In this case, these massless fermions of
definite chirality may correspond to the fermion spectrum as described by the SM.

To facilitate our discussion on possible routes to the SM, note that $SO(6)$
[$SO(3)$] is isomorphic to $SU(4)$ [$SU(2)$]. We can thus replace the
$\mathbf{Z}$-fields ($\mathbf{z}$-fields) with $SU(4)$-gauged [$SU(2)$-gauged]
ones in the above $\mathcal{M}_{3+1}$ effective theory. Now the Dirac fields
are the representations of $SU(4)$ and of $SU(2)$ simultaneously, represented
by $\psi_{v}^{\eta}=\binom{Q_{\mathbf{v}}^{\eta}}{l^{\eta}}$, where $\eta=1,2$
is $SU(2)$-valued and $v=1,2,3,4$ $SU(4)$-valued, e.g., $\psi_{\mathbf{v}%
}^{\eta}=Q_{\mathbf{v}}^{\eta}$ (quarks of three colors indexed by
$\mathbf{v}$), $\psi_{4}^{\eta}=l^{\eta}$ (leptons). Here, the lepton number
in our GIT works also as the fourth color, similarly to the Pati-Salam model
\cite{PatiS}. In the $\Gamma_{\mathrm{6D}}^{\mathrm{five}}$-diagonal basis,
the $SO(6)$ generators split into two $4\times4$ block-diagonal matrices.
Thus, $\psi_{v}^{\eta}$ for given $\eta$ has two four-component spinors, each
of which corresponds either to the four-dimensional complex representation of
$SU(4)$\ or to its conjugate \cite{group-book}.

Note that the index $\eta$ doubles the total number of elementary fermions,
which however is halved by the restriction of definite $\mathcal{M}_{9+1}$
chirality in our theory. This leaves the total number of elementary fermions
invariant. Now each family of matter fermions is labeled by one of the three
$SU(2)$ gauge fields (as well as the corresponding Higgs field) after symmetry
breaking of the $SU(2)$ symmetry. The SM symmetry $SU(3)\times SU(2)\times
U(1)$ can be obtained if symmetry breaking $SU(4)\rightarrow SU(3)\times
U(1)$\ occurs in our theory. The resulting SM is informationally incomplete
and effectively has $16-(8+3+1)=4$ matter fermions dark to the $SU(3)\times
SU(2)\times U(1)$ gauge fields. The SM symmetry might of course arise when
$\mathcal{M}_{6}$ is a coset space; for details see, e.g., \cite{KK-rev}.

\textit{Conclusions and discussions}.---To summarize, insisting on the
information-complete trinary description of nature, we have described very
briefly a unified theory for all forces and matter. Seen from our discussion,
the following two facts are strictly related or even mutually explained,
namely, (1) matter fermions have three families, each having 16 independent
matter fermions, and (2) spacetime is $(9+1)$-dimensional and displays
entanglement-driven symmetry breaking from $SO(9,1)$ down to $SO(3,1)\times
SO(6)$. The latter fact in turn explains why the observed spacetime is
$\mathcal{M}_{3+1}$ and the gauge group for matter fermions is $SO(6)$. The
information-complete trinary description of our theory naturally arises, thus
demonstrating that the overall picture is consistent with our previous
insistence on the information-completeness. Indeed, the ICP, which is essential for a new quantum formalism without the measurement postulate \cite{ICQT,icQFT}, works here as
\textit{the} fundamental principle restricting not only the required gauge
group, but more, such as the dimensions of spacetime and the number of
elementary matter fermions. Due to the discreteness of spacetime acting as a
natural regulator \cite{Thiemann-reg} and severe limitation of information by
the information-complete trinary description, our theory should be free of
singularities. Assuming $\mathcal{M}_{9+1}$ and the ICP, the theory seems to
be unique.

So far, our GIT predicts, in certain sense, some basic facts that are used in
the SM without any fundamental explanations. These include the dimensions of
spacetime, the three-family pattern, the number of elementary matter fermions,
and the possible origin of the Higgs mechanism. No previously known
\textit{single} theory could \textquotedblleft predict\textquotedblright\ all
these facts simultaneously. According to our GIT there is one and only one
matter fermion (a right-handed neutrino) that remains to be discovered for
each family. If the three-family structure of the cosmological constant [see
Eq.~(\ref{cos6d})] survives, it would be very exciting to see its cosmological test.

Is it possible to introduce supersymmetry/superstring in an
information-complete formalism? As supersymmetry predicts too many new
particles that are hard to suit the ICP, the answer to the question seems to
be gloomy. If this is indeed the case, the ICP could be a strong reason to
exclude supersymmetry. In future work, we need to consider the details of
quantum compactification and particularly, the Higgs physics to see if other
details of the SM could be derived within the GIT. Currently, we can only say
that the outputs of the GIT, as given here, looks very encouraging.


\begin{thebibliography}{99}                                                                                               %


\bibitem {Thiemann}T. Thiemann, \textit{Lectures on loop quantum gravity},
Lecture Notes in Physics \textbf{631}, 41-135 (2003).

\bibitem {Rovelli-book}C. Rovelli, \textit{Quantum Gravity} (Cambridge Univ.
Press, Cambridge, 2004).

\bibitem {loop25}C. Rovelli, \textit{Loop quantum gravity: the first twenty
five years}, Class. Quantum Grav. \textbf{28}, 153002 (2011).

\bibitem {Chiou}D.-W. Chiou, \textit{Loop quantum gravity}, Int. J. Mod. Phys.
D \textbf{24}, 1530005 (2015).

\bibitem {Smolin-PT}L. Smolin, \textit{Time, laws, and the future of
cosmology}, Phys. Today \textbf{67}, 38-43 (March, 2014).

\bibitem {Smolin-unify}L. Smolin, \textit{Unification of the state with the
dynamical law}, arXiv:1201.2632.

\bibitem {ICQT}Z.-B. Chen, \textit{The informationally-complete quantum
theory}, arXiv:1412.1079.

\bibitem {icQFT}Z.-B. Chen, \textit{An informationally-complete unification of
quantum spacetime and matter}, arXiv:1412.3662.

\bibitem {KK-rev}D. Bailin and A. Love, \textit{Kaluza-Klein theories}, Rep.
Prog. Phys. \textbf{50}, 1087-1170 (1987).

\bibitem {Kane-book}G. Kane, \textit{Modern Elementary Particle Physics:
Quarks, Leptons, and Their Interactions} (Addison-Wesley, Redwood City, CA, 1987).

\bibitem {su5}H. Georgi and S.L. Glashow, \textit{Unity of All
Elementary-Particle Forces}, Phys. Rev. Lett. \textbf{32}, 438-441 (1974).

\bibitem {so10-g}H. Georgi, \textit{The state of the art--gauge theories}, AIP
Conf. Proc. \textbf{23}, 575-582 (1975).

\bibitem {so10}H. Fritzsch and P. Minkowski, \textit{Unified interactions of
leptons and hadrons}, Ann. Phys. \textbf{93}, 193-266 (1975).

\bibitem {Witten}E. Witten, \textit{Fermion quantum numbers in Kaluza-Klein
theory}, in \textit{Proceedings of the 1983 Shelter Island Conference on
Quantum Field Theory and the Foundations of Physics}, eds. N. Khuri \textit{et
al}. (MIT Press, Cambridge, MA, 1985) 227-277.

\bibitem {Cho}Y.M. Cho, \textit{Higher-dimensional unifications of gravitation
and gauge theories}, J. Math. Phys. \textbf{16}, 2029-2035 (1975).

\bibitem {Manton}N.S. Manton, \textit{A new six-dimensional approach to the
Weinberg-Salam model}, Nucl. Phys. B \textbf{158}, 141-153 (1979); D.B.
Fairlie, \textit{Higgs' fields and the determination of the Weinberg angle},
Phys. Lett. B \textbf{82}, 97-100 (1979); P. Forg\'{a}cs and N.S. Manton,
\textit{Space-time symmetries in gauge theories}, Commun. Math. Phys.
\textbf{72}, 15-35 (1980).

\bibitem {Manton-chiral}N.S. Manton, Fermions and parity violation in
dimensional reduction schemes, Nucl. Phys. B \textbf{193}, 502-516 (1981).

\bibitem {Cianfrani}F. Cianfrani and G. Montani, \textit{Towards Loop Quantum
Gravity without the Time Gauge}, Phys. Rev. Lett. \textbf{102}, 091301 (2009);
\textit{Gravity in the presence of fermions as an SU(2) gauge theory}, Phys.
Rev. D \textbf{81}, 044015 (2010).

\bibitem {Salam}S. Randjbar-Daemi, A. Salam, and J.A. Strathdee,
\textit{Instability of higher dimensional Yang-Mills systems}, Phys. Lett B
\textbf{124}, 345-348 (1983).

\bibitem {origin}G. Dvali, S. Randjbar-Daemi, and R. Tabbash, \textit{The
origin of spontaneous symmetry breaking in theories with large extra
dimensions}, Phys. Rev. D \textbf{65}, 064021 (2002).

\bibitem {Hosotani}Y. Hosotani, \textit{Dynamical gauge-symmetry breaking and
left-right asymmetry in higher-dimensional theories}, Phys. Rev. D
\textbf{29}, 731-738 (1984).

\bibitem {PatiS}J.C. Pati and A. Salam, \textit{Lepton number as the fourth
\textquotedblleft color\textquotedblright}, Phys. Rev. D \textbf{10}, 275-289 (1974).

\bibitem {group-book}P. Ramond, \textit{Group Theory: A Physicist's Survey}
(Cambridge University Press, Cambridge, 2010).

\bibitem {Thiemann-reg}T. Thiemann, \textit{Quantum spin dynamics (QSD): V.
Quantum gravity as the natural regulator of the Hamiltonian constraint of
matter quantum field theories}, Class. Quantum Grav. \textbf{15}, 1281-1314 (1998).
\end{thebibliography}
\end{document}